\documentclass[journal]{IEEEtran}

\usepackage[utf8]{inputenc}
\usepackage[T1]{fontenc}
\usepackage{microtype}
\usepackage{amsmath,amssymb}
\usepackage{graphicx}
\usepackage{url}
\usepackage{hyperref}
\usepackage{xcolor}
\usepackage{orcidlink}

\hypersetup{
    colorlinks=true,
    linkcolor=black,
    citecolor=black,
    urlcolor=blue,
    pdftitle={A Field Guide to Decision Making},
    pdfauthor={Richard B. Arthur}
}

\begin{document}

% ----------  Title block  ----------
\title{A Field Guide to Decision Making}

\author{Richard~B.~Arthur\,\orcidlink{0000-0003-2960-9891}%
\thanks{Richard B. Arthur is with GE Aerospace Research, Niskayuna, NY 12309, USA (e-mail: arthurr@geaerospace.com).}%
\thanks{Digital Object Identifier 10.1109/MCSE.2026.3676981}%
\thanks{1521-9615 \copyright\ 2026 IEEE. All rights reserved, including rights for text and data mining, and training of artificial intelligence and similar technologies.}
}

\markboth{Computing in Science \& Engineering,~Vol.~X, No.~X, Month--Month~2026}%
{Arthur: A Field Guide to Decision Making}

\maketitle

% ----------  Abstract  ----------
\begin{abstract}
High-consequence decision making demands peak performance from individuals in positions of responsibility. Such executive authority bears the obligation to act despite uncertainty, limited resources, time constraints, and accountability risks. Tools and strategies to motivate confidence and foster risk tolerance must confront informational noise and can provide qualified accountability. Machine intelligence augments human cognition and perception to improve situational awareness, decision framing, flexibility, and coherence through agentic stewardship of contextual metadata. We examine systemic and behavioral factors crucial to address in scenarios encumbered by complexity, uncertainty, and urgency.
\end{abstract}


% ====================================================================
\section*{\hfill}
\vspace{-2.5em}
\IEEEPARstart{D}{ecisions} drive our actions in the present and shape the unfolding future. We make those decisions within the limits of our perception of the present moment and knowledge learned from past experiences.

In daily life, decision making can seem straightforward because we typically have sufficient time to consider alternatives, flexibility to change with minimal consequence, and the option to defer to commit or decline altogether. Professionals making consequential decisions bear the burden of \emph{accountability to act wisely} within the confines of windows of opportunity, available resources, and a tolerance for uncertainty. Decisions may incorporate tacit experience, psychological and emotional factors, and other contextual aspects that are difficult to distill from the human mind into machine data. Cassie Kozyrkov studies these interactions in her materials on decision intelligence~\cite{kozyrkov}.

To simplify the distinction of decisions that are consequential, consider whether the decision is documented into record, rather than requiring thresholds of impact, irreversibility, or dedication of resources. The recording serves to clarify action in the present and archive information for future reference. Two example contexts will illustrate use of decision records: 1)~medical patient records (diagnostic tests, therapies) and 2)~engineering records (design changes, tradeoff assessments).

% ====================================================================
\section{Decisions in Captivity}
Temporal orientation serves to frame decision scenarios: investigating the past, awareness in the present, and planning for the future. This section examines the current state of practice and highlights opportunities for improved information infrastructure.

\subsection{Present Orientation}
Decision-making authority is the defining responsibility for executive roles. That responsibility requires calculated commitment to action within an expected timeframe, budget, and availability of resources to develop sufficient confidence despite uncertainty and risk. Success results from balancing these components wisely, with a pragmatic perspective on current and future implications. Behaviors can be highly influenced by performance metrics, such as efficiency, throughput, yield, cycle time, and margin.

Agile methodologies motivate action through rapid launch-and-learn of minimally viable solutions, which subsequently improve through iteration. However, urgency to act and incentives to select expedient, cheap, or low-effort options can result in short-term, makeshift outcomes. Engineers refer to these quasi-solutions as \emph{tech debt}, a liability on a notional ledger of deferred resolutions, which have the potential for compounding risks, costs, and complexity in the intervening time.

Some fields sensitive to decision-making consistency have developed systems for disciplined compliance with procedural norms. Professionals that routinely perform high-stakes procedures, such as aviators, surgeons, and contract lawyers, employ formalized techniques for due diligence and confirmatory checklists. Regulatory authorities may also assert information be recorded for availability to audit.

Information technology (IT) capabilities can address operational encumbrances of urgency, complexity, and uncertainty. The concept of the digital thread~\cite{digitaltwin} represents a notional formalization of a connected genealogy of related records and data made accessible and interoperable for cross-functional collaboration: for example, linking engineering information and data from the design, manufacture, operation, and servicing of a product.

Knowledge management tools and practices facilitate capture, collaboration, and learning through annotation metadata, taxonomies for semantic alignment, and ontologies for understanding. Historically, knowledge management suffers from burdensome complexity and intrusive capture of tacit knowledge within otherwise intuitive workflows. Advances in artificial intelligence (AI), such as speech-to-text and large language models (LLMs), offer potential solutions for low-friction collection and curation of contextual metadata through side channels.

\subsection{Past Orientation}
Root cause analysis (RCA) is a well-established approach for responding to emergent, unexpected issues, such as a patient suffering from a rash or a pattern of part failures in a product. Questions posed consider evidence in the present relative to past conditions, events, decisions, and actions---such as: \emph{What was the cause? How and why did this happen?} and \emph{What was the cause of that cause?}---iteratively tracking contributing factors to identify root causes, perhaps formalized through Bayesian inference models.

The primary intent of RCA is to understand the scope of the problem, discover and validate causes, and then devise and evaluate options for a corrective solution. However, additional questions can offer crucial learning opportunities, such as: \emph{What decisions led to this? How and why were those decisions made?} (In particular, contextual framing of the decision at the time: alternatives, selection criteria, assumptions, unknowns).

Corrective urgency leading to a fix sufficient to move on, but incomplete as a remedy, can introduce tech debt. RCA encumbered with motives to assign blame may result in learned behaviors detrimental to promoting a culture of decision-making confidence and accountability.

\subsection{Future Orientation}
Anyone in a position to make consequential decisions will themselves bear repercussions for decisions made (or inaction) by way of being entrusted with that authority. Awareness of this accountability can influence behaviors in reflecting upon the possibility of judgment in hindsight, such as in the findings of a future RCA.

Rational strategies to mitigate that potential risk include consultation and conservatism. While measured implementation of those practices is reasonable and customary, particularly in regulated and high-consequence contexts, excessive deliberation, such as exhaustive analyses or consultations, can waste time, resources, and miss windows of opportunity to be effective. Likewise, compensating for uncertainty while under duress of urgency or inadequate resourcing promotes conservatism.

For example, consider product design ``overengineering,'' resulting in extraneous costs, delays, and reduction to performance specs or operational guidance. In health care, medical practitioners employ guarded caution with which patient medical records must be recorded in learned response to the proliferation of malpractice litigation and exposure to subpoena. This pragmatic hedge characterizes one of the tactics employed in the practice of ``defensive medicine'' to safeguard medical practitioners, often to the detriment of patients. Adverse consequences might include diminished quality of care, burden of increased tests (incremental logistics, costs, and time delays), and reduced candor in communication between physician and patient.

Awareness that high-consequence decisions may later be judged retrospectively can encourage behavioral flaws toward irresponsible caution~\cite{irresponsiblecaution}.

% ====================================================================
\section{Decisions in the Wild}
The executive decision makers entrusted with responsibility for high-consequence decisions confront systemic and behavioral factors that undermine confidence. The gathering and assessment of supporting information must contend with volatility, uncertainty, complexity, and ambiguity (VUCA). Decision urgency can further constrain time and availability of resources to achieve desired clarity, resulting in satisficing behaviors described by Herbert Simon's \emph{Bounded Rationality}~\cite{simon}.

Additionally, organizational boundaries and intervals of elapsed time introduce opportunities for inconsistency between related decisions. The politics of blame and socialized negativity bias expose decisions and actions to selective retrospective judgment in hindsight.

Unmitigated, these risks undermine confidence, motivate a culture of mediocrity through blame-avoidant behaviors, and result in wasted time, resources, funds, diminishing operational robustness, agility, and eroding trust required to coordinate. Systemic, behavioral, and exogenous factors can also impede responsible action, particularly under duress and urgency when events place lives, property, and order in jeopardy.

\subsection{VUCA Fog of War}
The U.S. Army War College characterized the post-Cold War operational landscape with the term VUCA: for volatility, uncertainty, complexity, and ambiguity, elements of the ``fog of war'' in military doctrine stretching back to Sun Tzu.

The VUCA terms may be defined~\cite{vuca}:
\begin{itemize}
    \item \emph{volatility}: prevalent forces/catalysts of rapid change
    \item \emph{uncertainty}: factors limiting prediction/confidence
    \item \emph{complexity}: many-factor interdependencies
    \item \emph{ambiguity}: unclear, nuanced, and mixed interpretations.
\end{itemize}

While urgent decision making can employ systematic preparedness to mitigate complexity and uncertainty, volatility and ambiguity pose more elusive challenges.

\emph{Complexity} may be addressed through preparedness, learning from past experiences, codifying choices, and processes where best practices as standard operating procedures (SOP) can bound and simplify options and context. These proactive measures help reduce errors and improve timely, sound, and consistent decisions, even under the duress of urgency. SOPs are commonplace in both medical practices and product manufacturing to address routine tasks and reinforce compliance. Increasingly powerful analytical tools aim to tame ever-growing complexity, most notably through the recent advances deriving models from vast data through machine learning (ML), including LLM semantic models.

\emph{Uncertainty} and error persist in limiting data-derived model performance. While gathering additional data can reduce epistemic uncertainty, aleatoric uncertainty arising from noisy, dimensionally sparse, and difficult (or expensive) to measure data persist in confounding predictive capability. Furthermore, the principle of ``garbage in, garbage out'' can instead become ``garbage amplified'' with na\"ive acceptance of, or reliance upon, insufficiently informed (poorly trained) models.

\emph{Volatility} can rapidly erode even well-established situational clarity, as critical factors become obsolete and consequently require persistent adaptation to changes, guided by appropriate guardrails~\cite{guardrails}.

Although SOPs, standard work, best practices, and data-derived models (ML) can provide preparedness strategies ahead of or in swift response to scenarios driven by urgency, model applicability will remain confined to the time period and scope of their framing reference data.

Some information and data remain unknowable until their emergence in future events, confounding even extensive precautions. Only persistent vigilance over the passing of time may afford desired confidence, including resolution of the shifting interpretations that confer \emph{ambiguity}. Fluid interpretations of ambiguity form one source of perceived inconsistency; additional factors can also challenge coherence of thought across decisions.

\subsection{Inconsistency}
Decisions are rarely isolated or independent. Interrelated decisions may occur sequenced over time and across stakeholders with significantly different experience, perspectives, and incentives. Affected parties and outside observers correspondingly form interpretations through their own distinct lenses. Perceived inconsistency among these participants undermines confidence and trust.

IT tools and practices can compensate for the unreliability of the human mind to reliably perform precise and accurate recall from memory or to apply persistent focused attention over extended spans of time. Traditional approaches, such as shared database archives and procedural checklists, facilitate consistency over time and between collaborators.

For example, at an annual physical checkup, a physician will consult medical records to refresh their memory of that patient's health context. The discussion should retain coherence of thought and strategy for care despite the physician having seen numerous other patients during the intervening year. The records capture and summarize contextually relevant information and changes outside the examination, including family medical history, visits to other specialists, and perhaps insurance or regulatory policies for approval of diagnostics or therapies. As medical institutions impose patient throughput targets on practitioners, timely and accurate records become crucial for physicians to maintain the urgent pace of visits while minimizing poor outcomes due to error or omission. Therefore, patient record systems must be sufficiently robust and intuitive to use to maintain consistency event to event and among practice specializations.

Discontinuities due to organizational boundaries, roles with differing incentives, mismatched access to resources, or from agents with disparate sociopolitical norms can introduce inconsistencies in values and assumptions influencing consequential decision making. The impact of such factors can range from na\"ive misalignment to fomenting competition that motivates behaviors counter to desired institutional results.

For example, engineering design inherently navigates between tradeoffs allocating costs and resources versus targets, such as performance, efficiency, reliability, and time to market. Consider installation of a freeway ramp with bounds on project cost and urgency to complete. Savings on upfront capital costs, time, and resources for the project may lead to higher total costs and effort for ongoing maintenance. Prioritization criteria dependencies are sensitive to budget cycles, access to resources, and situational variables across municipal, contractor, and regulatory decision makers. Misaligned or contradictory assumptions shift the organizational burden needed to address consequent tech debt.

Inconsistency undermines trust and clarity, which can result in avoidance of accountability when responsibility becomes instead perceived as liability.

\subsection{Self-Interest}
Decisions of consequence routinely prompt unsolicited observations, commentary, and actions by others. This can be praise or admiration, lending \emph{credit}, but results short of flawless success can provoke motivated scrutiny, critique, and judgment, subjecting the decision maker to \emph{blame}.

Oxford University findings published by Christopher Hood find blame carries four times the cultural weight as comparable credit~\cite{hood}. Social amplification of blame, termed \emph{negativity bias}, reinforces this effect, with successes taken for granted while failures are reliably and asymmetrically spotlighted. Anticipation of exposure to liability and other retrospective judgment in hindsight motivates devoting greater effort to avoiding blame than to earning credit. Hood terms diminished potential resulting from blame-avoidant behaviors \emph{mediocrity bias}.

Examples of blame-avoidance strategies Hood cites are:
\begin{itemize}
    \item \emph{Deflect}: Scapegoat, delegate, encode as policy.
    \item \emph{Distract}: ``Spin,'' bury in news/procedural cycle.
    \item \emph{Decline}: Shun risk of toxic exposure despite merit.
    \item \emph{Diffuse}: Mutual deniability (credit optional).
\end{itemize}

Self-interest behaviors and survival instincts conflate challenges already prevalent in the dynamic, inscrutable, interconnected, and nuanced VUCA environment. Passive forms of accountability evasion include reluctance to document decisions and actions: For example, to curtail defensively the candor, transparency, and completeness of information supplied to systems of record (e.g., cautious detail in patient medical records, prioritizing mitigation of risk from a malpractice subpoena over patient care utility).

Less passive tactics may cross lines established on principles of integrity, accountability, and even legality. The present political landscape has highlighted intentional disinformation as one blatant method to evade and reassign accountability. Its merited discussion lies outside of the scope of this article, however, beyond acknowledging the practice as a factor to consider.

Game theory offers a framework for characterizing and analyzing interaction behaviors among interdependent decision makers. Urgent and high-consequence scenarios amplify the stakes in calculations for compromise, prioritization, and bounds for inclusion. Measures should be taken to promote trust and confidence, and to recognize and compensate for gullibility, insincerity, and ineptitude.

Example archetypes characterizing extremes of behaviors and motivations are:
\begin{itemize}
    \item \emph{Wise}: Acknowledges uncertainty and their own ignorance, proactively and thoroughly records accurate data with transparent intent to qualify, quantify, and learn.
    \item \emph{Benignly na\"ive}: Lacks awareness, experience, and/or resources, and is open to learn if afforded the opportunity.
    \item \emph{Stubbornly na\"ive}: Lacks awareness and interest to inquire or improve, firmly clutch unfounded certainty, and is subdued by cognitive biases.
    \item \emph{Defensive}: Reflexively avoids accountability by employing blame avoidance strategies by default, seeking survival as best-case outcome after the storm passes.
    \item \emph{Self-interested}: Prioritizes personal objectives over organizational goals, willfully projects unmerited certainty, exploits logical fallacies, and counters inquiry aggressively.
\end{itemize}

Decisions ``in the wild'' cannot presume the luxury of peers exclusively from \emph{the wise}. On the other hand, draconian distrust and verification anticipating purely \emph{selfish} coactors may waste time, resources, and foment distrust, prompting \emph{defensive} postures among coactors.

Confidence lies at the foundation of high-consequence decision making. Insufficient confidence and coordination confronting VUCA and coherent connectedness over time and across stakeholders can prompt flawed decision-making behaviors, from unintentional inconsistency to avoidance of accountability. Technological advances in knowledge systems offer tactics to improve collaborations across organizations and over time, improve consistency of thought, and operationalize adaptation.

% ====================================================================
\section{Decisions Evolved}
Enlisting the rapidly advancing capabilities of digital infrastructure and machine intelligence, we can envision a collaborative knowledge system to bolster confidence and accountability with improved contextual clarity of the past, awareness in the present, and anticipated preparedness for future needs. Implementing such a framework requires cultural commitment from leadership and an integrated knowledge infrastructure to navigate decision making within the VUCA environment.

We set the following goals to improve decision making:
\begin{itemize}
    \item confidence to act despite uncertainty
    \item decision coherence and consistency
    \item preparedness through adaptation by design.
\end{itemize}

\subsection{Decision Provenance}
The effectiveness of a technical knowledge infrastructure to achieve these goals despite VUCA factors will require leadership and cultural commitment to safeguard \emph{learning} as a strategic operational foundation. Learning shifts perspective from decisions as potential liabilities to viewing them as key control points for adaptation and improvement.

First, recognize decisions as the most consequential result of the myriad activities gathering, processing, qualifying, quantifying, verifying, validating, analyzing, synthesizing, modeling, simulating, evaluating, and framing data and information. Second, notionally transform decisions from a result to be archived to become a primary index for underlying data, models, and analyses supporting said decision. Finally, anticipate cause to revisit a decision in the future, formally recording decision-time caveats, concerns, assumptions, limitations, and unknowns as searchable decision metadata.

These metadata constitute \emph{decision provenance}, providing transparency into the framing context at the time the decision was made. The term adopts the notion of provenance as metadata as used by scientists to improve understanding and reproducibility of experimental data and analyses~\cite{prov}. Decision provenance metadata provide similar pedigree for the origin and lineage of decisions of significance.

\emph{Decision provenance} (metadata) examples are:
\begin{itemize}
    \item criteria to select and evaluate
    \item alternatives considered
    \item assumptions
    \item constraints (budget, time, expertise, \dots)
    \item (known) unknowns
    \item supporting references \dots
\end{itemize}

Documenting present constraints and tacit factors seen as relevant to making the decision can mitigate concern that a future retrospective will miss present nuance or limitations, which may no longer hold in that future. An example is awareness of additional diagnostic test options for a patient, but election to hold off based on logistical and financial constraints.

Numerous prior efforts to capture implicit, tacit knowledge have failed due to challenges, ranging from management of semantic consistency and complexity to onerous, nuanced, and resistant human interaction. Advances in AI technologies have shifted the state of the possible for both the computational obstacles (through data-derived models, such as ML and LLMs) and user acceptance of collaboration with intelligent systems (normalized through home automation and chatbots).

While legacy processes may already capture some of this information informally in annotations of archived reference documents (e.g., slides, spreadsheets, or e-mail), the explicit mapping of contextual metadata into a searchable knowledge representation forms a foundational function of the envisioned knowledge infrastructure.

\subsection{To Act Despite Uncertainty}
This archival of decisions, their supporting data, and provenance metadata into an enterprise system for knowledge stewardship establishes a recognized authority that can be queried to clarify sensitivities and limitations current to decision making. Cultural leadership to recognize and employ an authoritative reference offers decision makers \emph{qualified} accountability as a safeguard against capricious blame or liability.

Assurance of qualified accountability can lend confidence to commit \emph{to act despite uncertainty} and reduce deliberation and conservatism, even when constrained by urgency: for example, present shortage of the otherwise most desirable raw material for manufacturing a part.

In the design of a system for capturing decision context (with candor and clarity), the provenance metadata merit information sensitivity assessment akin to the underlying data and decisions themselves. Exposing the thought processes and state of mind framing consequential decisions, the assumptions, unknowns, evaluation criteria, and alternatives taken together merit careful assessment for confidentiality and protection.

Additionally, the adaptive and dynamic nature of a system for knowledge stewardship creates opportunities to undermine integrity if unethical actors can rewrite history, motivated by stealing credit or avoiding blame. Without reliable and robust version controls and immutable audit records, the system will struggle to establish itself as an authoritative system record and cornerstone for providing qualified accountability. Blockchain techniques may offer the needed trust, transparency, and auditability.

\subsection{Coherence and Consistency}
\emph{The Minding Organization} describes problem-solving strategies to envision the (desirable) future and bring it into the present~\cite{minding}. This facility to shift perspective through time and across organizations with a coherent frame of reference develops a \emph{continuum mindset}. This perspective pursues coherent (consistently logical) thought spanning intervals of time and crossing organizational discontinuities, to deliver precocious enterprise situational awareness and to robustly adapt to emerging knowledge.

Records of decision provenance support a coherent frame of reference from which to understand, learn, and adapt over time, offering decision makers in the present the opportunity to act with the advantage of future insights. This capability can uncannily mediate the effects of VUCA that foster inconsistency, wasteful deliberation, and mediocrity resulting from hedging against accountability.

Implemented with a robust ontology or elegant language model, the process of submitting decision provenance can detect inconsistencies between previously archived assumptions, criteria, and identified unknowns, flagging contradiction to the attention of both the present and prior decision makers. This promotes improved coherence between stakeholders and over time: For instance, when prescribing a newly available drug with milder side effects, identifying prior patients under similar treatment who may now benefit from switching.

Future reassessment of decisions may then be carried out with the benefit of leveraging relevant prior efforts that considered figures of merit, candidate alternatives, processes to reproduce supporting analyses and previous results, mappings between dependent decisions, and even further cross-linking via like assumptions and unknowns. An example is prior effort selecting part suppliers and performance metrics to apply in contract management.

\subsection{Adaptation by Design}
Digital systems customarily perform searches against \emph{archived} data \emph{retrospectively}. But affordable abundance of processing in modern digital technology offers opportunity to build far more sophisticated information infrastructure.

Entwining decisions with contextual provenance enables language-savvy software agents to perform directed semantic searches on unstructured data feeds to find information valuable for clarification and consistency. An example is intelligent agents that \emph{persistently} search \emph{emerging} data, then react in a prescribed manner to changes (\emph{volatility}), developments and discoveries (\emph{uncertainty}), and clarifications (\emph{ambiguity}).

This machinery offers the capability to perform \emph{adaptation by design}. For example, the system for enterprise knowledge stewardship may employ an agentic framework to periodically or continually monitor sensitive criteria, assumptions, or unknowns against diverse information feeds. When a contradiction, confirmation, or discovery is detected, an agent can alert relevant stakeholders, indicating specific decision dependencies.

Looking from the vantage of the decision in the past, this agent performs a \emph{search into the future} to allow correction of a flawed assumption and revision of related decisions. This capability empowers consequential decisions to be made with greater confidence and reduced time, effort, and conservatism imposed by the prevailing uncertainties of the past: for instance, commitment to approve a product design on the condition that the product is never installed above a certain altitude.

In effect, the system acts as a safety net to hedge upon a decision \emph{conditionally}, aware that a priori recorded concerns of consequence (\emph{known unknowns} or asserted \emph{assumptions}) will be persistently monitored and trigger prescribed review upon confirmation (or contradiction).

The integration of high-consequence decisions and contextual provenance into a semantics-capable agentic system offers a tremendous opportunity toward an \emph{awakened enterprise}~\cite{awakened}. For example, such a system would be capable of executing other novel forms of risk mitigation, such as monitoring a catastrophic ``black swan'' scenario by submitting a contingent antidecision, encoding into the provenance failure modes, or similar factors meriting vigilance in rare but high-risk scenarios meriting proactive lead time for adaptive response.

% ====================================================================
\section{Conclusion}
Systematized data gathering and knowledge management underlie the ``awakened enterprise'' strategy to collect, manage, and monitor decision provenance to tame VUCA complexity and pursue agility over volatility, understanding over uncertainty, and judgment over ambiguity. The visibility and consistency of decision context over spans of time and across organizational boundaries affords decision makers' \emph{qualified accountability} and ability to act despite uncertainty and opportunistically adapt to emergent changes in the future. Timely, proactive intervention reduces overmanagement and mitigates fear of retrospective liability. This approach promotes the candor and transparency for effective organizational operations and learning.

% ====================================================================

% ----------  Author biography  ----------
\vspace{1em}
\noindent\textbf{RICHARD B. ARTHUR} serves as a senior principal engineer at GE Aerospace, Niskayuna, NY, 12309, USA. His research interests include computational methods, knowledge systems, and model-based engineering. Contact him at \href{mailto:arthurr@geaerospace.com}{arthurr@geaerospace.com}.

\end{document}